\documentclass{ws-ijmpe}

\usepackage{epsfig}
\usepackage{amsmath}
\usepackage{sidecap}

\usepackage{color}

\definecolor{grey}{rgb}{0.93,0.93,0.93}
\definecolor{lred}{rgb}{1,0.4,0.4}

\begin{document}

\markboth{N. Schindzielorz, J. Erler, P. Kl\"upfel, P.--G. Reinhard,
  G. Hager}
{Fission of super-heavy nuclei explored with Skyrme forces}

%
\catchline{}{}{}{}{}
%

\title{Fission of super-heavy nuclei explored with Skyrme forces}

\author{N. Schindzielorz, J. Erler, P. Kl\"upfel, P.--G. Reinhard
}
\address{\it
Institut f\"ur Theoretische Physik II, Universit\"at Erlangen, D-91058,
Erlangen, Germany}

\author{G. Hager}
\address{\it
Regionales Rechenzentrum der  Universit\"at Erlangen, D-91058,
Erlangen, Germany\\
}

\maketitle

\begin{history}
\received{(received date)} \revised{(revised date)}
\end{history}

\begin{abstract}
We present a large scale survey of life-times for spontaneous fission
in the regime of super-heavy elements (SHE), i.e. nuclei with
Z=104-122.  This is done on the basis of the Skyrme-Hartree-Fock
model.  The axially symmetric fission path is computed using a
quadrupole constraint.  Self-consistent cranking is used for the
collective masses and associated quantum corrections.  The actual
tunneling probability is estimated by the WKB approximation.  Three
typical Skyrme forces are used to explore the sensitivity of the
results.  Benchmarks in the regime Z=104-108 show an acceptable
agreement.  The general systematics reflects nicely the islands of
shell stabilization and the crossover from $\alpha$-decay to  
fission for the decay chains from the region of Z/N=118/176.. 
\end{abstract}

\maketitle

\section{Introduction}

Self-consistent mean-field (SCMF) models have proven to be reliable
tools for describing nuclear structure and dynamics, for reviews see
\cite{Ben03aR,Vre05aR,Stone_rew}.  They are steadily developing to
improve the quality of the predictions and to accommodate more and
more observables.  One of the very demanding observables are fission
life-times.  Fission barriers have been discussed already in the early
stages of SCMF models and even been used as benchmark for calibration
\cite{skms}.  Fission life-times are much more involved as their
computation requires not only the potential energy surface along the
fission path, but also the corresponding collective masses and a safe
estimate of the collective ground state correlations for the initial
state.  There are thus not so many self-consistent calculations of
fission life-times\cite{Ber01a,War06} -- 
{mostly using still approximate masses } { and
quantum corrections (=zero point energies \cite{Rei87aR})}.
The vast majority of calculations employs the microscopic-macroscopic
method which combines shell corrections with a macroscopic liquid-drop
model background, see e.g.  \cite{Mol87a,Smo95a}.  It is the aim of
this contribution to explore the systematics of fission life-times all
over the landscape of super-heavy elements.  This is done using the
Skyrme-Hartree-Fock method as one widely used nuclear SCMF model
\cite{Ben03aR}.

The paper is outlined as follows: In section \ref{sec:formal} we
briefly review the theoretical background and explain how we compute
fission life-times.  Results are presented in section
\ref{sec:results} covering tests with experimentally known life-times,
comparison with $\alpha$-decay life-times and a large scale
systematics.

\section{Formal framework}
\label{sec:formal}

The starting point for the self-consistent microscopic description
is the SHF energy functional 
\begin{equation}
  E
  =
  E_\mathrm{Skyrme}(\rho,\tau,\mathbf{J};\mathbf{j},
  \mbox{\boldmath$\sigma$},\mbox{\boldmath$\tau$};\chi)
  \quad
\end{equation}
which is expressed in terms of a few local densities and currents
obtained as sums over single-particle wave functions: density $\rho$,
kinetic density $\tau$, spin-orbit density ${\bf J}$, current ${\bf
j}$, spin density ${\bf\sigma}$, and pair density $\chi$ where each
occurs twice, once for protons and once for neutrons.  The zero-range
pairing functional is handled in a stabilized BCS ansatz to achieve 
smooth transitions of the occupation numbers. For details see \cite{Erl08}.
There exist various parameterizations for the Skyrme functional.
In order to explore the possible sensitivity of the fission life-times
to the parameterization, we 
{confine the survey}
to three 
sufficiently different parameterizations:
SkP as a force with effective nucleon mass $m^*/m=1$
\cite{skp}, SkI3 as a fit which 
{has very low mass $m^*/m=0.6$}
and maps the relativistic iso-vector
structure of the spin-orbit force \cite{ski3}, and Sly6 as a fit which 
{has $m^*/m=0.7$}
and
includes information on isotopic trends and neutron matter
\cite{sly46}.

The computation of the fission path and the ingredients of the
corresponding collective Hamiltonian is detailed in
\cite{Fle05a,Klu08a}. We give a brief summary. The mean-field
equations are derived variationally from the given energy
functional. They are complemented by a quadrupole constraint to
generate the fission path thus reading
\begin{subequations}
\begin{eqnarray}
  &&
  \big[\hat{h}-\lambda\hat{Q}_{20}\big]|\Phi_q\rangle
  =
  \mathcal{E}|\Phi_q\rangle
  \quad,\hspace{0.5cm} q =\langle\Phi_q|\hat{Q}_{20}|\Phi_q\rangle
\\
  &&
  \mathcal{V}(q)
  =
  E_\mathrm{Skyrme}(\rho_q,\tau_q,
                    \mathbf{J}_q,\chi_q)
  \quad,
\label{eq:rawpot}
\end{eqnarray}
\end{subequations}
where $\hat{h}$ is the mean-field Hamiltonian and $\rho_q$
is the local density for the state $|\Phi_{q}\rangle$, similarly for
$\tau_q$, $\mathbf{J}_q$, and
$\chi_q$.  The optimal fission path should, in fact, be
generated by adiabatic time-dependent Hartree-Fock
\cite{Rei87aR,Ska08a}.  The quadrupole constraint is a plausible
and generally used approximation to that.
To compute the collective mass along the fission path, we need to
explore the dynamical response of the system to changing deformation,
commonly called self-consistent cranking.
To that end we use the collective-momentum operator $\hat{P}_{20}$
(and its collective momentum p) as
additional constraint. The  $\hat{P}_{20}$ is deduced as generator
of deformation, the response to $\hat{P}_{20}$ creates a
momentum-dependent path and subsequently total energy, from which
the mass is finally obtained as second order term in collective momentum,
altogether
\begin{subequations}
\begin{eqnarray}
  &&
  \hat{P}_{20}:\qquad
  \hat{P}_{20}|\Phi_q\rangle
  \propto
  \mathrm{i}\partial_q|\Phi_q\rangle
  \quad,
\\
  &&
  \big[\hat{h}-\lambda\hat{Q}_{20}-\mu\hat{P}_{20}\big]
  |\Phi_{qp}\rangle
  =
  \mathcal{E}|\Phi_{qp}\rangle
  \quad,
\\
  &&
  B_{20}
  =
  \frac{1}{2}\partial_{p_{20}}^2E(\rho_{qp},...)\big|_{p=0}
  \quad.
\end{eqnarray}
\end{subequations}
Analogously, the momentum of inertia $\Theta_x$ for rotations about
the $x$- and $y$-axis is computed by self-consistent cranking.
The potential $\mathcal{V}$ as given in eq. (\ref{eq:rawpot}) has to be
augmented by quantum corrections to account for angular-momentum
projection and spurious zero-point quadrupole motion \cite{Rei87aR}
yielding the net potential
\begin{eqnarray}
  \mathcal{V}
  \quad\longrightarrow\quad
  V
  &=&
  \mathcal{V}
  -
  B_{20}\langle\hat{P}_{20}^2\rangle
  -
  \frac{1}{8\langle\hat{P}_{20}^2\rangle}\partial_q^2\mathcal{V}
  -
  \frac{\langle\hat{J}_x^2+\hat{J}_y^2\rangle}{2\Theta_x}
  \quad.
\label{eq:zpecor}
\end{eqnarray}

Having determined the properly corrected collective potential
$V(\alpha_{20})$ and inverse collective mass $B(q)$, we
compute the fission life-time in the standard semi-classical fashion
using the WKB expressions for tunneling probability $W$ and repetition
time $T$:
\begin{eqnarray}
  W
  =
  \exp{\Big(\!-2\!\int_b^cdq\sqrt{\frac{V(q)\!-\!E}{B_{20}}}\Big)}
  \quad,\quad
\label{equ:period-koll}
  T
  &=&
  \hbar\int_a^b\!dq\frac{1}{\sqrt{B_{20}(E\!-\!V(q))}}
  \quad.
\end{eqnarray}
The integrals are evaluated by trapezoidal rule with separate handling
at the divergent $\sqrt{E-V}$ at the classical turning points.  A
quantity which enters very critically these expressions is the
tunneling energy E, which is also the ground state energy of the
nucleus before fission.  That is computed with great care in a fully
quantum-mechanical treatment of the collective quadrupole oscillations
using all five quadrupole degrees-of-freedom, for details of that part
of the calculations see \cite{Fle05a,Klu08a}.

{
It is to be noted that the fission path employs only axially symmetric
shapes.  One knows from actinides that triaxial shapes can produce
barriers which are 0--2~MeV lower.  The present results are thus to be
understood as an upper limit on barriers and
lifetimes.
Actinides show the famous double-humped barrier.  The outer barrier
dissappears for SHE leaving only one, the formerly inner barrier
\cite{Bur04}. This barrier is related to still symmetric shapes while
asymmetry develops quickly when stepping down further towards fission.
The fact that we have only one barrier simplifies the survey of
fission properties of SHE.
}

\section{Results and discussion}
\label{sec:results}

\begin{SCfigure}[0.5]
\hspace*{0.5cm}{\epsfig{file=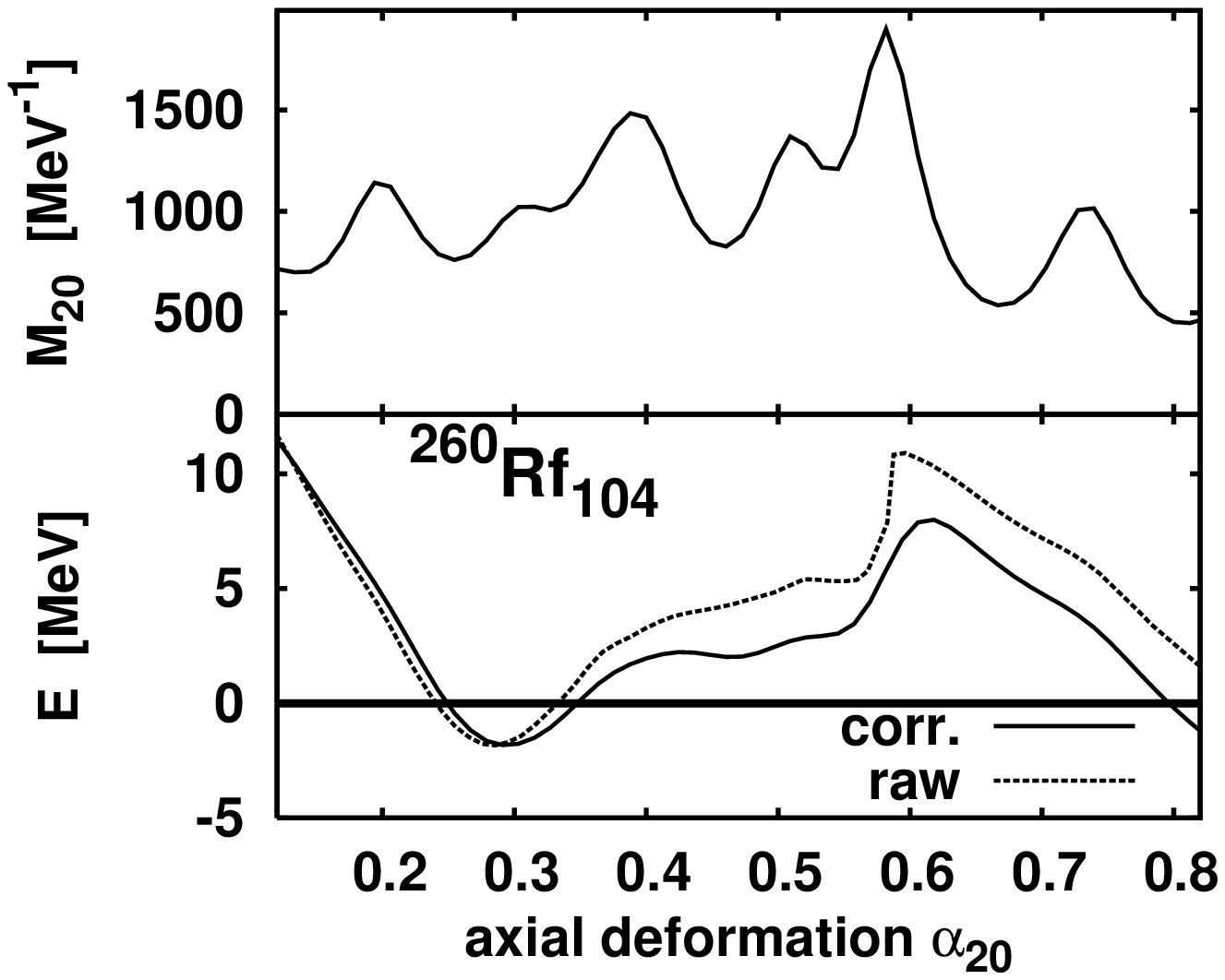,width=6.5cm}}
\caption{\label{fig:barriers-mass2-bw}
Example of potential energy (lower) and collective mass
(upper) along the axially symmetric fission path of 
{$^{260}$Rf}
.
For the potential energy, we show the raw expectation value
$\mathcal{V}$ and the corrected one $V$, see eq.~(\ref{eq:zpecor}).
}
\end{SCfigure}
Fig.~\ref{fig:barriers-mass2-bw} shows collective potential and mass
along the fission path of 
{$^{260}$Rf}
. The quantum corrections on the
potential (lower panel) reduce the barrier by an important amount
making 3 to 6 orders lower life-times than would be obtained with
the raw potential \cite{Rei87aR}. 
{
One recognizes a steep increase of the energy at deformation
$\alpha_{20}\approx 0.6$ connected with a quick change of
hexadecapole momentum $\alpha_{40}$. The system moves from
one ridge (with high  $\alpha_{40}$) to another ridge
(with low  $\alpha_{40}$). The segregation of the fission landscape
into ridges is well known for actinides, see e.g. \cite{Moe01a},
still visible for that smaller SHE, and much smoothened with
further increasing system size.
}
The quadrupole mass (upper panel)
shows large fluctuations. High values are related to regions of
level crossings. The detailed pattern are quantitatively important and
can hardly be simulated by some constant collective mass.

\begin{SCfigure}[0.5]
{\epsfig{file=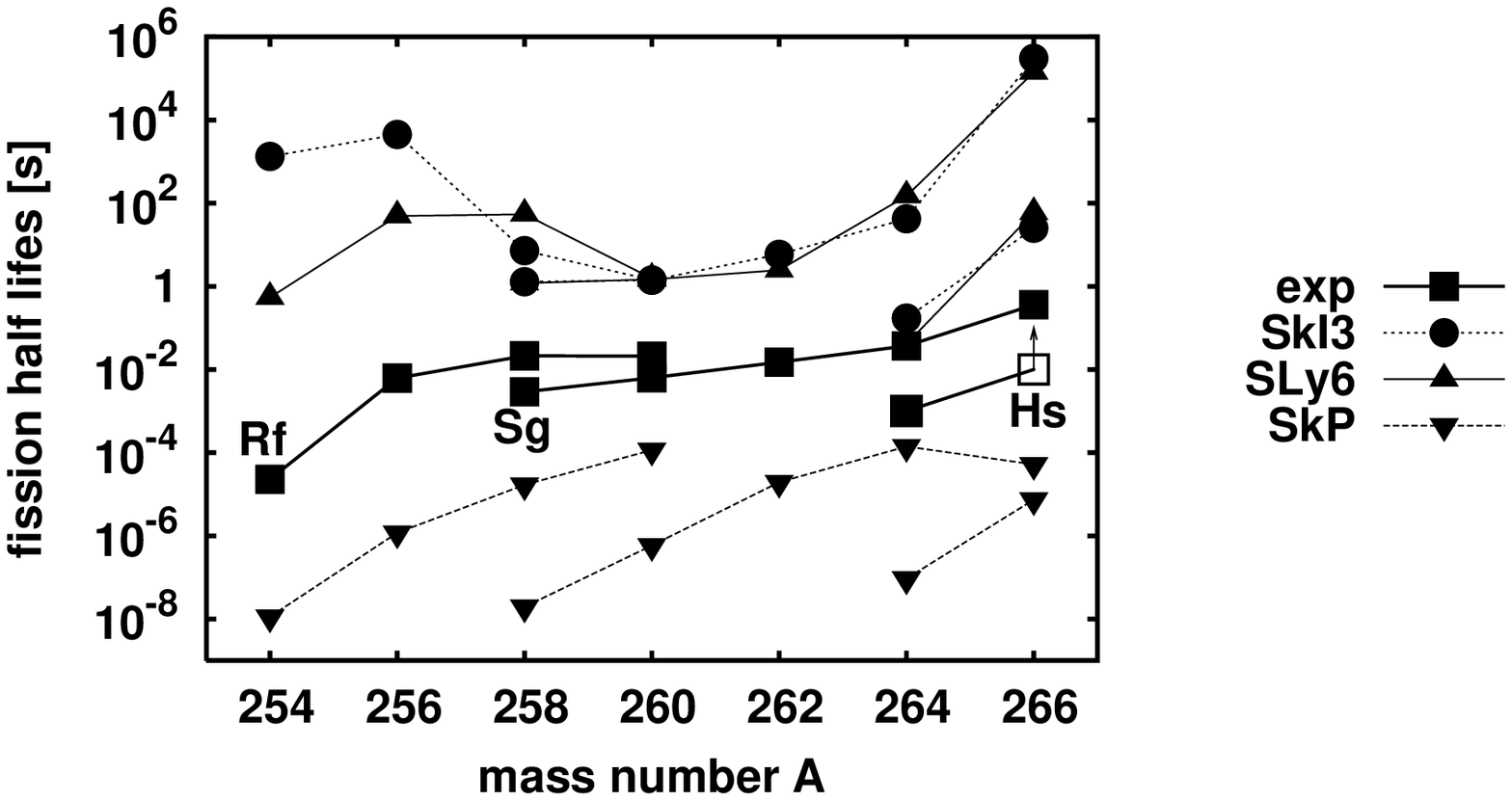,width=9.5cm}}
\caption{\label{fig:compare-half-lives-2-bw}
Fission life-times computed with three different Skyrme
parameterizations, as indicated, and compared with
data from 
\protect\cite{expRf254,expRf256,expRf258,expSg262,expSg266,expHs266}.
}
\end{SCfigure}
As a benchmark, we compare in fig.~\ref{fig:compare-half-lives-2-bw}
with experimental results available at the lower edge of
SHE.  There
are large differences in the predictions from the various forces. But
the experimental values are nicely bracketed by the band of
predictions and the trends within each force compare well with
experiment. The deviations of $\pm 4$ orders of magnitude look
huge. But they seem bearable in view of the extremely delicate balance
in computing the tunneling rates.

\begin{figure}
\centerline{\epsfig{file=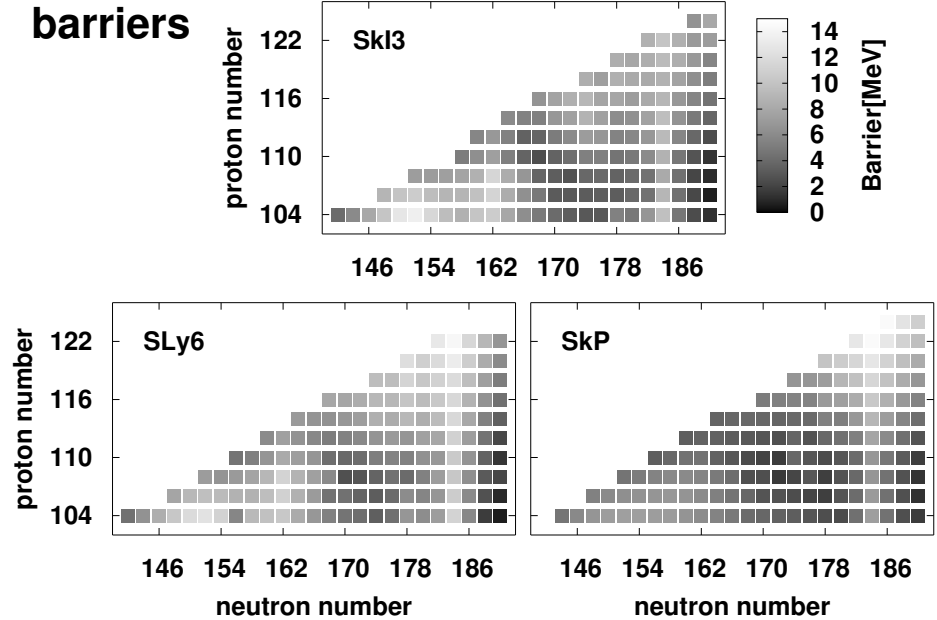,width=\linewidth}}
\caption{\label{fig:system-barriers-bw}
Fission barriers for a broad variety of super-heavy elements
and for three Skyrme parameterizations as indicated.
}
\end{figure}
\begin{figure}
\centerline{\epsfig{file=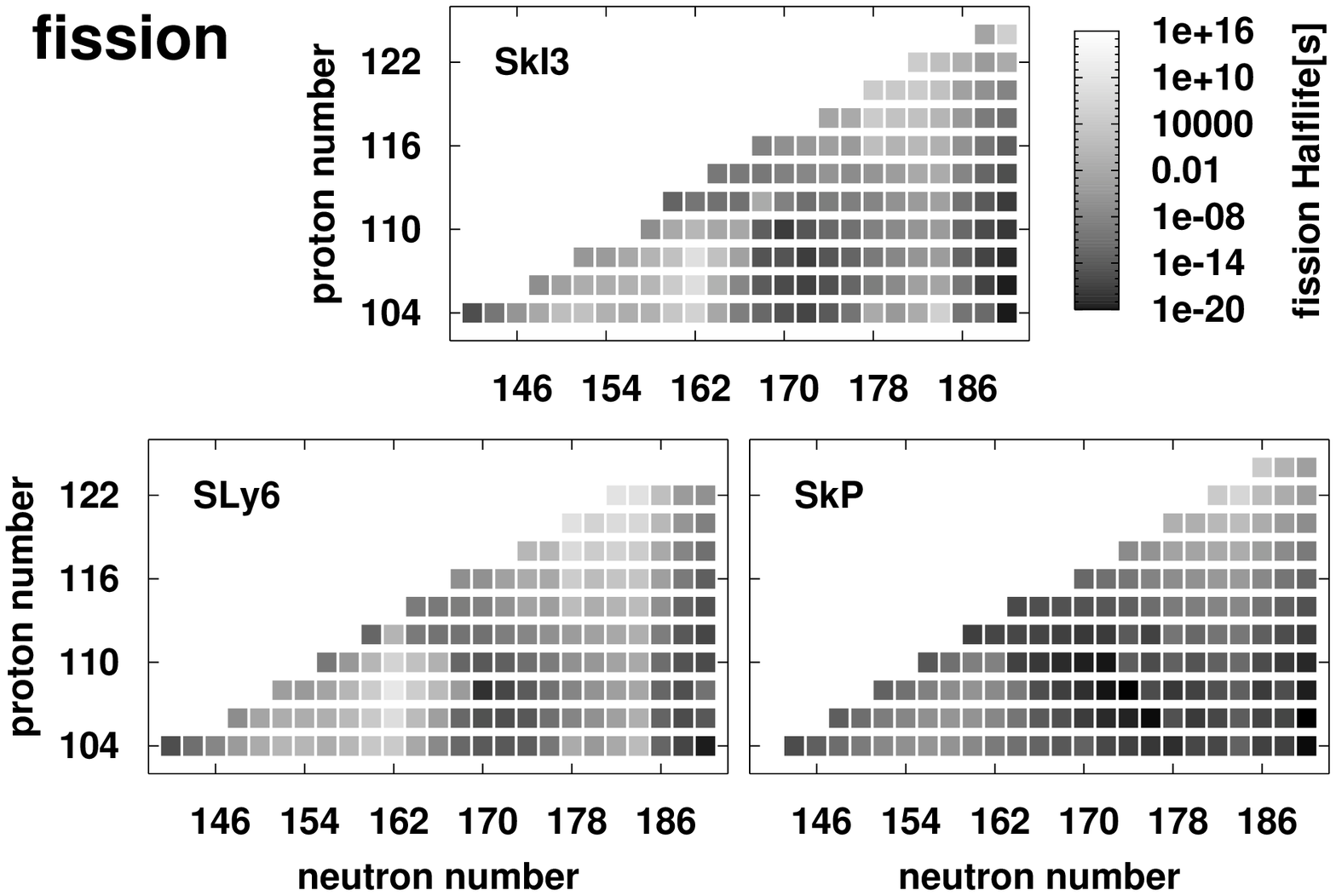,width=\linewidth}}
\caption{\label{fig:system-lifetimes-bw}
Fission life-times
for a broad variety of super-heavy elements
and for three Skyrme parameterizations as indicated.
}
\end{figure}
We are now going to present systematics of results all over the
landscape of SHE where we consider all elements which are found to be
stable against immediate nucleon emission at ground state and along
the whole fission path. Fig.~\ref{fig:system-barriers-bw} shows the
fission barriers. {They agree} in value and systematics with 
{the previous survey of barriers \cite{Bur04}}.
One sees two islands of enhanced fission 
stability (high barriers), one of deformed SHE around Z/N=104/152 and the other
of spherical SHE around 120/184. The magic neutron number N=184 is
also clearly marked while no unambiguous sign of a magic proton number
can be found in that upper island \cite{Ben01a}.  All three forces
agree in predicting a ridge of very low barriers between the upper and
the lower stability island. 
{Barriers however, are mainly indications of stability. 
The full information on path, {mass, and quantum corrections}
is required to calculate life-times.}

Fig.~\ref{fig:system-lifetimes-bw} shows the corresponding fission
life-times. The difference in barrier heights from 0 to 12 MeV
translates to life-times from almost immediate decay to $10^{16}$ s,
demonstrating again the enormous sensitivity of fission life-times to
any ingredient in its computation. The long lived SHE are found in the
two islands of stability. Practically immediate decay appears in the
ridge of instability between the islands. The variance of the
predictions is moderate in the islands (see also
fig.~\ref{fig:compare-half-lives-2-bw}) and significantly larger in
the unstable region. The latter region is not only unstable but also
very hard to control theoretically.

\begin{figure}
\centerline{\epsfig{file=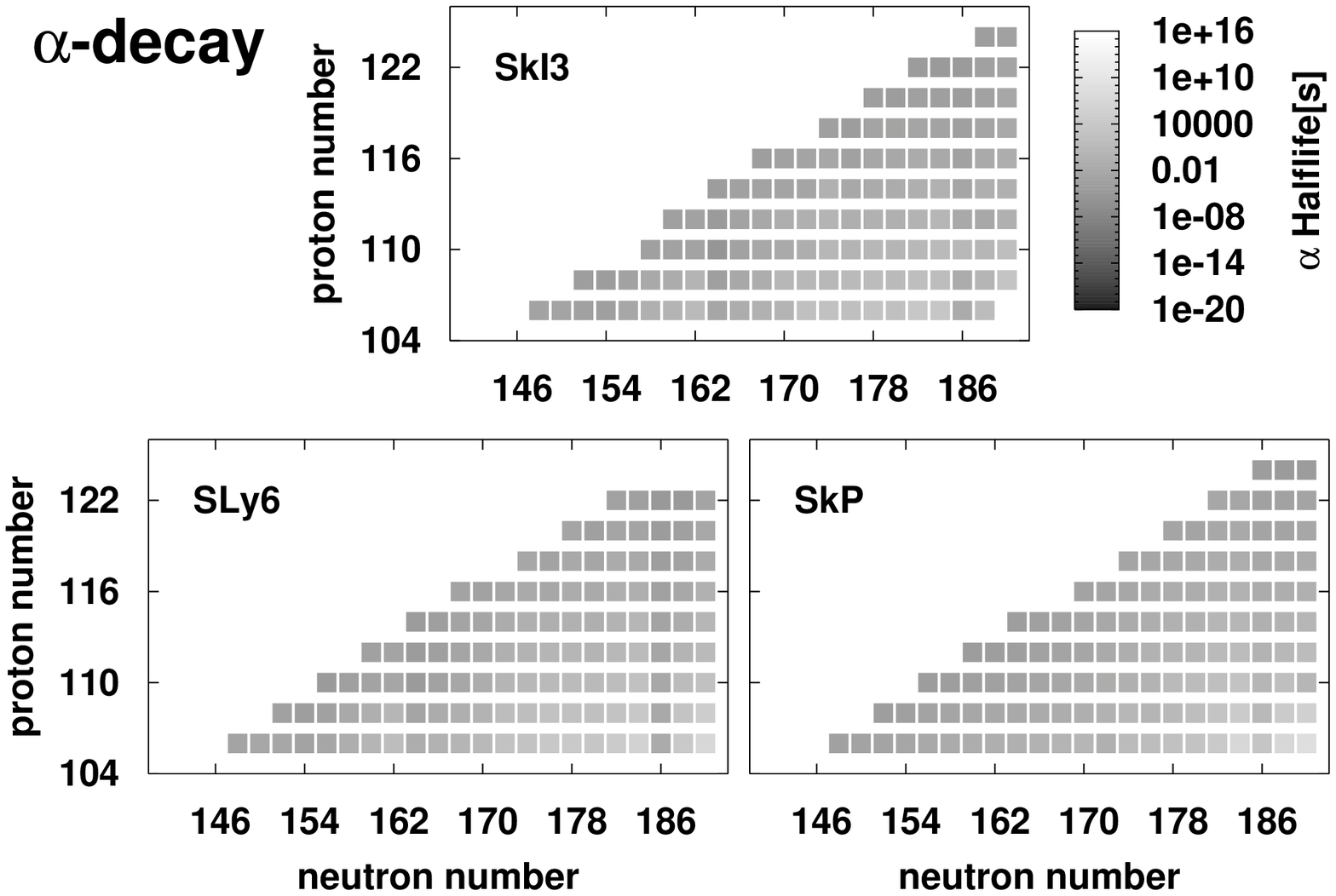,width=\linewidth}}
\caption{\label{fig:system-alphalifetimes-bw}
$\alpha$-decay lifetime
for a broad variety of super-heavy elements
and for three Skyrme parameterizations as indicated.
}
\end{figure}
The competing decay channel for many SHE is $\alpha$-decay. We have
computed the $\alpha$ life-times from the $Q_\alpha$ reaction energies
using a recently improved recipe \cite{Sam07} based on the Viola
systematics.  Fig.~\ref{fig:system-alphalifetimes-bw} shows the results. The
$\alpha$ life-times change much more smoothly. There are no shell
effects visible and the variance of life-times all over the landscape
is very much smaller. Comparing with
fig.~\ref{fig:system-lifetimes-bw}, one sees that $\alpha$-decay is
indeed the dominating decay channel in the islands of
stability. Stepping down from the upper island, one sees a crossover
of $\alpha$-decay to spontaneous fission at about Z=112 or 110, much
in accordance with the experimental findings so far, see
e.g. \cite{exp116}.

%
The relation between $Q_\alpha$ value and $\alpha$ life-time is
comparatively simple 
{when adopting the Viola systematics \cite{Sam07}}.
It is interesting to ponder whether one could
hope for a similarly simple direct connection between fission barrier
and fission life-time.  A comparison of figure
\ref{fig:system-barriers-bw} with \ref{fig:system-lifetimes-bw} shows
that the very gross trends may look similar. 
{
We have checked whether one could establish a simple relation between
barriers and lifetimes. This turned out to be impossible.
}
Fission life-times are very subtle quantities which depend on 
{more ingredients than just barriers alone.}

\begin{SCfigure}[0.5]
{\epsfig{file=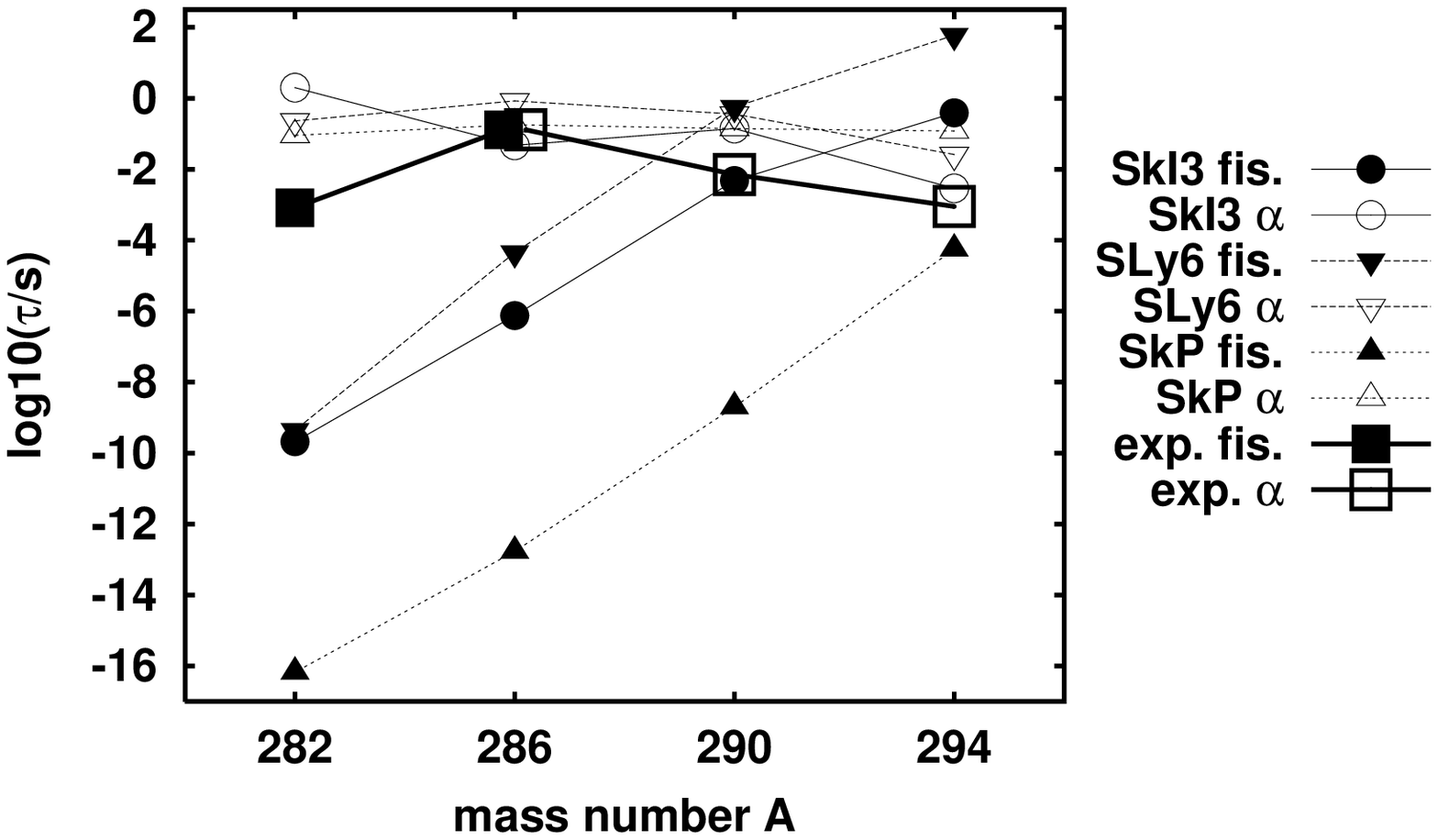,width=7.5cm}}
\caption{\label{fig:trends-halflifes-bw}
Fission life-times and $\alpha$-decay life-times 
for $\quad$
 elements along the decay chain from $Z/N=118/176$
computed with three different Skyrme
parameterizations, as indicated, and compared with experimental
data from \protect\cite{exp118}.
}
\end{SCfigure}
There exist a few fission life-times along the decay channel from the
upper island of stability. Fig.~\ref{fig:trends-halflifes-bw} compares
computed life-times with data for fission and $\alpha$-decay. The
$\alpha$ life-times are well reproduced by the calculations, but the
fission life-times are dramatically underestimated towards the lower
end of the chain. The reason for that is still unclear. The mismatch
calls for further investigations.

\section{Conclusions}

Fission life-times have been computed using the Skyrme-Hartree-Fock
method.  The fission path has been generated with a quadrupole
constraint where only axially symmetric deformations are considered.
For the corresponding collective mass, we use self-consistent
cranking.  The quantum corrections to the collective potential
(angular momentum projection, vibrational zero-point energy) are
properly taken into account.  The fission life-time is computed for
thus given potential and mass by the WKB approximation, while the
ground state energy, which is at the same time the entrance energy for
fission, is computed quantum mechanically.  Results have been produced
for three different Skyrme forces, SkP, SLy6 and SkI3, to explore the
sensitivity to the parameterization.  For comparison, we have also
computed the $\alpha$-decay life-times using the Viola systematics.

A first test was performed by comparing with known fission life-times
in the lower region of super-heavy elements, Z=104-108.  The
theoretical results gather around the experimental values with
deviations of $\pm 2$ orders of magnitude.  That can be called a
promising agreement, the more so, as the isotropic trends are well
reproduced.  The second test went for fission and $\alpha$-decay
life-times in the $\alpha$-decay chain from 118/176.  The
$\alpha$-decay life-times are well reproduced.  But it turned out that
the fission life-times are grossly underestimated by all three forces
towards the lower end of the chain at 112/170.  The reasons for that
mismatch are not yet clear.  It is to be reminded that these
super-heavy nuclei have extremely soft deformation energy surfaces
with several shape isomers which complicates the determination of the
ground state.
Notwithstanding that open problem, we have produced a systematic
survey of fission barriers, fission life-times and $\alpha$-decay
life-times for all conceivable super-heavy elements in the range
Z=104-124 and N=150-190.  Barriers and fission life-times show nicely
the islands of shell stabilization around 104/152 and 120/184.  All
forces predict a band of fission instability being orthogonal to the
line connecting the two islands and crossing that around
112/166. There seems to be no path which could connect the 120/184
region with the lower island while avoiding fission.
While the fission life-times show dramatic variation over the chart of
super-heavy elements (from instability to 10$^{16}$s), the
$\alpha$-decay times vary gently with small overall changes and
without visible shell effects.  The general crossover from
$\alpha$-decay to fission along the decay chains from the upper island
is qualitatively reproduced by all three forces in the survey.

Altogether, the results are promising and challenging at the same
time.  They call for further investigations exploring more
systematically the sensitivity to the Skyrme parameterization and
improving the description in the very soft transitional region around
112/166.

\section*{Acknowledgments}

We thank the regional computing center of the university
Erlangen-N\"urnberg for generous supply of computer time for the
demanding calculations.  The work was supported by the BMBF under
contracts 06 ER 808.


\begin{thebibliography}{99}



\bibitem{Ben03aR} 
  M. Bender, P.-H. Heenen, and P.-G. Reinhard,
  {\it Rev. Mod. Rhys.} {\bf 75}, 121 (2003).
\bibitem{Vre05aR}
D. Vretenar, A.V. Afanasjev, G.A. Lalazissis, P. Ring,
{\it Phys. Rep.},
{\bf 409}, 101 (2005).
\bibitem{Stone_rew} 
  J.R. Stone and P.-G. Reinhard,
  {\it Prog. Part. Nucl. Phys.} {\bf 58}, 587 (2007).
\bibitem{skms}
  J. Bartel,
  P. Quentin, M. Brack, C. Guet, and H.-B. H\aa{a}kansson,
  {\it Nucl. Phys.} A{\bf 386}, 79 (1982).

\bibitem{Ber01a}
J.--F. Berger, L. Bitaud, J. Decharge, M. Girod, and K. Dietrich,
{\it Nucl. Phys.} A{\bf 685}, 1c (2001).


\bibitem{War06}
  M. Warda {\it et al},
  {\it Phys. Scr. T}, \textbf{125}, 226 (2006).

\bibitem{Rei87aR}
P.-G. Reinhard and K. Goeke,
{\it Rep. Prog. Phys.} {\bf{50}}, 1 (1987).


\bibitem{Mol87a}
{P. M{\"o}ller, J. R. Nix, and W. J. Swiatecki}, 
{\it Nucl. Phys.} A{\bf 469}, 1 (1987).


\bibitem{Smo95a}
R. Smola{\'n}czuk, J. Skalski, and A. Sobiczewski, 
  {\it Phys. Rev. C} {\bf 52}, 1871 (1995).


\bibitem{Erl08}
{J. Erler, P. Kl\"upfel, and P.--G. Reinhard},
{\it Eur. Phys. J. A},
{\bf 37},81 (2008)


\bibitem{skp}
{J. Dobaczewski, H. Flocard, and J. Treiner}, 
  {\it Nucl. Phys.} A{\bf 422}, 103 (1984).


\bibitem{ski3}
  P.-G. Reinhard and H. Flocard,
  {\it Nucl. Phys.} A{\bf 584}, 467 (1995).


\bibitem{sly46}
  E. Chabanat, P. Bonche, P. Haensel, J. Meyer, and R. Schaeffer,
  {\it Nucl. Phys.} A627, 710 (1997).




\bibitem{Fle05a}
P. Fleischer, P. Kl\"upfel, P.--G. Reinhard, and J. A. Maruhn,
{\it Phys. Rev. C},
{\bf 70}, 054321 (2004)


\bibitem{Klu08a}
{P. Kl\"upfel, J. Erler, P.--G. Reinhard, and J. A. Maruhn},
{\it Eur. Phys. J. A},
{\bf 37},343 (2008)




\bibitem{Ska08a}
J. Skalski,
{\it Phys. Rev. C} {\bf 77}, 064610 (2008).



\bibitem{Bur04}
  T. B{\"u}rvenich {\it et al},
  {\it Phys. Rev. C}, \textbf{69}, 014307 (2004).

\bibitem{Moe01a}
P. M\"oller {\it et al},
{\it Nature} {\bf 409}, 785 (2001).


\bibitem{exp116}
  Yu. Ts. Oganessian {\it et al},
  {\it Eur. Phys. J. A}, \textbf{15}, 201 (2002).

\bibitem{exp118}
  Yu. Ts. Oganessian {\it et al},
  {\it Phys. Rev. C}, \textbf{74}, 044602 (2006).

\bibitem{expRf254}
  G. Audi {\it et al},
  {\it Nucl. Phys. A}, \textbf{729}, 3 (2003).

\bibitem{expRf256}
  F. P. Hessberger {\it et al},
  {\it Z. Phys. A}, \textbf{359}, 415 (1997).

\bibitem{expRf258}
  J. M. Gates {\it et al},
  {\it Phys. Rev. C}, \textbf{77}, 034603 (2008).

\bibitem{expSg262}
  K. Gregorich {\it et al},
  {\it Phys. Rev. C}, \textbf{74}, 044611 (2006).

\bibitem{expSg266}
  J. Dvorak {\it et al},
  {\it Phys. Rev. Lett.}, \textbf{100}, 132503 (2008).

\bibitem{expHs266}
  S. Hofmann {\it et al},
  {\it Eur. Phys. J. A}, \textbf{10}, 5 (2001).




\bibitem{Ben01a}
M. Bender, W. Nazarewicz, and P.--G. Reinhard,
{\it  Phys. Lett. B },
{\bf 515}, 42 (2001).

\bibitem{Sam07}
{C. Samanta{,} P. Roy Chowdhury, and D.N. Basu},
{\it Nucl. Phys. A} {\bf 789}, 142 (2007).





\end{thebibliography}
\end{document}